\documentclass[twocolumn]{autart}   
\usepackage{graphicx}
\usepackage{natbib}
\usepackage[caption=false]{subfig}
\usepackage{epsfig}
\usepackage{epstopdf}
\usepackage{amsmath}
\usepackage{amssymb}
\usepackage{mathtools}
\usepackage{color}
\usepackage{enumerate}

\usepackage{comment}
\usepackage{stfloats}

\begin{document}

\begin{frontmatter}

\title{Minimum-Time Trajectory Optimization With \\ Data-Based Models: A Linear Programming Approach\thanksref{footnoteinfo}} 
\vspace{-2mm}

\thanks[footnoteinfo]{This paper was not presented at any IFAC meeting. Corresponding author N.~Li. Email: {\it nanli@auburn.edu}}

\author[AU]{Nan Li}\ead{nanli@auburn.edu},
\author[AU]{Ehsan Taheri}\ead{etaheri@auburn.edu},
\author[UM]{Ilya Kolmanovsky}\ead{ilya@umich.edu},
\author[TAMU]{Dimitar Filev}\ead{dfilev@tamu.edu}

\address[AU]{Department of Aerospace Engineering, Auburn University, Auburn, AL, USA}
\address[UM]{Department of Aerospace Engineering, University of Michigan, Ann Arbor, MI, USA}
\address[TAMU]{Hagler Institute for Advanced Study, Texas A\&M University, College Station, TX, USA}
\vspace{-2mm}

\begin{abstract}
In this paper, we develop a computationally-efficient approach to minimum-time trajectory optimization using input-output data-based models, to produce an end-to-end data-to-control solution to time-optimal planning/control of dynamic systems and hence facilitate their autonomous operation. The approach integrates a non-parametric data-based model for trajectory prediction and a continuous optimization formulation based on an exponential weighting scheme for minimum-time trajectory planning. The optimization problem in its final form is a linear program and is easy to solve. We validate the approach and illustrate its application with a spacecraft relative motion planning problem.
\end{abstract}
\vspace{-2mm}

\vspace{-2mm}

\end{frontmatter}

\section{Introduction}\label{sec:1}

Minimum-time trajectory optimization (also known as ``time-optimal control'') is frequently involved in robot path planning and tracking \citep{lepetivc2003time,verscheure2009time}, space mission design and operation \citep{shirazi2018spacecraft}, and other disciplines wherever a task is desired to be accomplished in a least amount of time, typically subject to limited resources. Due to its significant relevance, the investigation into this topic has been extensive over decades \citep{kalman1959general,lasalle1959time,o1981discrete}.

The formulation of a minimum-time trajectory optimization problem can be either in continuous time or in~discrete time. In continuous time, solution techniques are usually based on indirect methods (also~known as ``variational methods''): One first applies Pontryagin's maximum principle to reduce the trajectory optimization problem to a two-point boundary value problem~(TPBVP), and then solves the TPBVP using a~shooting method \citep{trelat2012optimal,taheri2017co}. In discrete time, in contrast, direct methods are more often considered. Various direct methods for minimum-time trajectory optimization in discrete time have been proposed in the literature: In the approach of \cite{carvallo1990milp}, a~minimum-time problem is reformulated into a mixed-integer program, where a set of Boolean variables are used to indicate if a~given target condition is reached at each time step over the planning horizon. In the approaches of \cite{van2011model} and \cite{zhang2014minimum}, a minimum-time problem is solved using a bi-level algorithm, where the lower level solves a fixed-horizon trajectory planning problem treating the target condition as a terminal constraint and the upper level adjusts the planning horizon of the lower-level problem to find the minimum horizon length such that the lower-level problem admits a feasible solution. In the approaches of \cite{rosmann2015timed} and \cite{wang2017minimum}, the time is scaled and the scaling factor is treated as an optimization variable. This way, minimizing the final time of the original free-final-time problem is achieved through minimizing the scaling factor in a~related fixed-final-time problem. However, in such an approach based on time scaling, the optimization problem is necessarily nonlinear and nonconvex, even for linear systems. In the approach of \cite{verschueren2017stabilizing}, the deviation from the target condition is penalized with a weight that increases exponentially with time. It is shown that a minimum-time trajectory solution can be obtained if the weight parameter is chosen to be sufficiently high. The approaches reviewed above typically use a state-space model of the considered dynamic system to predict the trajectories. A common approach to obtaining a state-space model, to be used for trajectory optimization, is to first derive the model from first principles and then calibrate its parameters according to prior knowledge and/or data of the system.

Recently, a non-parametric modeling approach based~on behavioral systems theory is gaining attention due to its unique applicability to emerging data-driven paradigms \citep{markovsky2021behavioral}. This approach uses input-output time-series data of a given system to build up a (non-parametric) predictive model, which is referred to as a {\it data-based model} in this paper, rather than using the data to identify/calibrate a (parametric) state-space model. The integration of such data-based models into predictive control algorithms has been investigated by various researchers, e.g., \cite{coulson2019data,coulson2021distributionally,berberich2020data,baros2022online,huang2023robust}, and has demonstrated superior performance than conventional control methods in a number of applications \citep{elokda2021data,huang2021decentralized,chinde2022data}. Along with bypassing the step of state-space model identification/calibration, such a predictive control algorithm using an input-output data-based model determines control input values directly based on output measurements and does not require state estimation using an observer. Therefore, it provides an ``end-to-end'' data-to-control solution, which may be simpler to implement from the point of view of practitioners and may facilitate fully autonomous operation of future intelligent systems.

In the above context, the goal of this paper is to develop an approach to minimum-time trajectory optimization based on input-output data-based models, to produce an end-to-end data-to-control solution to many time-optimal planning and control problems in robotics, aerospace, and other disciplines. To the best of our knowledge, there has been no previous work addressing this topic. We focus on linear time-invariant systems, for which we adopt the approach based on behavioral systems theory to build up a non-parametric data-based model for trajectory prediction. We then exploit an~exponential weighting scheme extended from the one of \cite{verschueren2017stabilizing} to solve for minimum-time trajectories. We show that the optimization problem in its final form is a linear program and hence is easy to solve. Main contributions of this paper are as follows:
\begin{itemize}
    \item We develop an approach to minimum-time trajectory optimization based on input-output data-based models, which is the first such approach in the literature.
    \item We extend the trajectory prediction method in previous predictive control algorithms using data-based models (e.g., the ones in \cite{coulson2019data,coulson2021distributionally,berberich2020data,baros2022online,huang2023robust}) to enable predicting trajectories over an~extended planning horizon without relying on a~high-dimensional model. This extension is especially relevant to trajectory optimization because a~long planning horizon is not rare in a trajectory optimization setting. Our method is inspired by the multiple shooting method \citep{trelat2012optimal}.
    \item Using an exponential weighting scheme extended from that of \cite{verschueren2017stabilizing}, we formulate the minimum-time trajectory optimization problem into a linear program, which is easy to solve. We prove that minimum-time trajectories can be obtained with the linear program if a weight parameter is chosen to be sufficiently high. In particular, we make the following extensions to the exponential weighting scheme: 1)~The scheme was used in \cite{verschueren2017stabilizing} for minimum-time trajectory planning based on a state-space model; we extend the scheme to apply it to minimum-time trajectory planning based on an input-output data-based model. 2) The scheme was used in \cite{verschueren2017stabilizing} for point-to-point trajectory planning; we extend the scheme to more general point-to-set trajectory planning. To enable these extensions, different assumptions are made, and, correspondingly, new proofs are developed to show the ability of this exponential weighting scheme to produce minimum-time trajectory solutions.
    \item We validate the developed approach to minimum-time trajectory optimization and illustrate its application with an aerospace example.
\end{itemize}

{\it Organization:} The minimum-time trajectory optimization problem addressed in this paper is described in~Section~\ref{sec:2}. We introduce the method using a non-parametric input-output data-based model to predict trajectories over an extended planning horizon in~Section~\ref{sec:3}. We reformulate the minimum-time trajectory optimization problem into a linear program based on an~exponential weighting scheme and prove its theoretical properties in~Section~\ref{sec:4}. A spacecraft relative motion planning problem is considered in Section~\ref{sec:5} to illustrate the approach. The paper is concluded in Section~\ref{sec:6}.

{\it Notations:} The symbol $\mathbb{R}$ denotes the set of real numbers and $\mathbb{Z}$ the set of integers; $\mathbb{R}^n$ denotes the set of $n$-dimensional real vectors, $\mathbb{R}^{n \times m}$ the set of $n$-by-$m$ real matrices, and $\mathbb{Z}_{\ge a}$ the set of integers that are greater than or equal to $a$; $I_n$ denotes the $n$-dimensional identity matrix, $0_{n,m}$ the $n$-by-$m$ zero matrix, and $1_n$ the $n$-dimensional column vector of ones. Given multiple vectors $v_k \in \mathbb{R}^{n_k}$ or matrices with the same number of columns $M_k \in \mathbb{R}^{n_k \times m}$, $k = 1,...,K$, the operator $\text{col}(\cdot)$ stack them on top of one another, i.e., $\text{col}(v_1, ..., v_K) = [v_1^{\top}, ..., v_K^{\top}]^{\top}$ and $\text{col}(M_1, ..., M_K) = [M_1^{\top}, ..., M_K^{\top}]^{\top}$. For a discrete-time signal $z(\cdot): \mathbb{Z} \to \mathbb{R}^n$, we use ${\bf z}_{[a:b]}$, with $a,b \in \mathbb{Z}$ and $a \le b$, to denote $\text{col}(z(a), ..., z(b))$. We call both the sequence $z(a), ..., z(b)$ and the column vector ${\bf z}_{[a:b]} = \text{col}(z(a), ..., z(b))$ a trajectory (of length $l = b-a+1$). 

\section{Problem Statement}\label{sec:2}

We study trajectory optimization problems associated with finite-dimensional linear time-invariant systems which can be represented in state-space form as
\begin{subequations}\label{equ:1}
\begin{align}
x(t+1) &= A x(t) + B u(t) \\
y(t) &= C x(t) + D u(t)
\end{align}
\end{subequations}
where $t \in \mathbb{Z}$ denotes the discrete time step, $x(t) \in \mathbb{R}^n$ represents the system state at time $t$, $u(t) \in \mathbb{R}^m$ is the control input, $y(t) \in \mathbb{R}^p$ is the output, and $A$, $B$, $C$ and $D$ are matrices of appropriate dimensions. We make the following assumption about the system:

{\it Assumption~1:} The system is controllable and observable.

Given a state-space model \eqref{equ:1} of the system, we can write the following equation that relates an input trajectory of length $l$, $l \in \mathbb{Z}_{\ge 1}$, to its corresponding output trajectory:
\begin{equation}\label{equ:2}
{\bf y}_{[0:l-1]} = \mathcal{O}_l x(0) + \mathcal{C}_l {\bf u}_{[0:l-1]}
\end{equation}
where $x(0)$ is the system state at the initial time of the trajectory, and $\mathcal{O}_l$ and $\mathcal{C}_l$ are matrices defined as follows:
\begin{equation}\label{equ:3}
\mathcal{O}_l = \begin{bmatrix} C \\
CA \\ \vdots \\ CA^{l-1} \end{bmatrix} \quad \mathcal{C}_l = \begin{bmatrix} D & & & \\
CB & D & & & \\
\vdots & \ddots & \ddots & \\
CA^{l-2}B & \cdots & CB & D \end{bmatrix}
\end{equation}
The smallest integer $l$ such that the matrix $\mathcal{O}_l$ defined above has full rank is called the {\it lag} of the system and denoted as $l_{\min}$. Under {\it Assumption~1}, $l_{\min}$ exists and satisfies $1 \le l_{\min} \le n$. 

Given an arbitrary pair of input trajectory ${\bf u}_{[0:l-1]}$ and output trajectory ${\bf y}_{[0:l-1]}$, both of length $l$, if \eqref{equ:2} holds for some $x(0) \in \mathbb{R}^n$, then the pair $({\bf u}_{[0:l-1]},{\bf y}_{[0:l-1]})$ is called {\it admissible} by the system. In particular, an admissible pair of input-output trajectories of length $l$, $({\bf u}_{[0:l-1]},{\bf y}_{[0:l-1]})$, with $l \ge l_{\min}$, corresponds to a unique initial state $x(0)$, which can be determined according to~\eqref{equ:2} as follows:
\begin{equation}\label{equ:4}
x(0) = \mathcal{O}_l^\dagger ({\bf y}_{[0:l-1]} - \mathcal{C}_l {\bf u}_{[0:l-1]})
\end{equation}
where $(\cdot)^\dagger$ denotes the Moore-Penrose pseudoinverse.

In this paper, we focus on minimum-time trajectory optimization problems given in the following form:
\begin{subequations}\label{equ:5}
\begin{align}
\min_{u(\cdot), y(\cdot), T \ge 0} &\quad T \label{equ:51} \\
\text{s.t.} &\quad ({\bf u}_{[-K_i:-1]},{\bf y}_{[-K_i:-1]}) = ({\bf u}_i,{\bf y}_i) \label{equ:52} \\
&\quad {\bf y}_{[T:T+K_f-1]} \in Y_f \label{equ:53} \\
&\quad c(u(t), y(t)) \le 0, \quad t = 0, ..., T+K_f-1 \label{equ:54}
\end{align}
\end{subequations}
where $K_i \ge l_{\min}$; ${\bf u}_i \in \mathbb{R}^{m K_i}$ and ${\bf y}_i \in \mathbb{R}^{p K_i}$ represent given initial conditions for the input and output trajectories; $K_f \ge 1$; $Y_f \subset \mathbb{R}^{p K_f}$ represents a target set for the output trajectory; and $c(u(t), y(t)) \le 0$ represents prescribed path constraints for the trajectory to satisfy. The goal represented by the cost function in \eqref{equ:51} and the constraint in \eqref{equ:53} is to drive the output trajectory to reach the target set $Y_f$ in the minimum time, starting with the initial condition in \eqref{equ:52}, while satisfying the path constraints in \eqref{equ:54}. We first make the following assumption about the initial condition $({\bf u}_i,{\bf y}_i)$:

{\it Assumption~2:} The pair $({\bf u}_i,{\bf y}_i)$ represents an admissible pair of input-output trajectories (of length $K_i$) that satisfies the path constraints $c(u(t), y(t)) \le 0$ at all times.

{\it Assumption~2} is reasonable because ${\bf u}_i$ and ${\bf y}_i$, according to \eqref{equ:52}, represent the input and output trajectories of the system over the past $K_i$ time steps, and hence, the pair $({\bf u}_i,{\bf y}_i)$ is supposed to be admissible and satisfies any path constraints. Under {\it Assumption~2}, given a state-space model \eqref{equ:1} of the system, \eqref{equ:52} corresponds to the following initial condition for the state:
\begin{equation}\label{equ:6}
x(0) = A^{K_i} \mathcal{O}_{K_i}^\dagger ({\bf y}_i - \mathcal{C}_{K_i} {\bf u}_i) + \begin{bmatrix} A^{K_i-1}B & \cdots & AB & B\end{bmatrix} {\bf u}_i
\end{equation}
where $\mathcal{O}_{K_i}$ and $\mathcal{C}_{K_i}$ are the matrices defined in \eqref{equ:3} with $l = K_i$.

We consider polyhedral target set $Y_f \subset \mathbb{R}^{p K_f}$ and linear-inequality path constraints $c(u(t), y(t)) \le 0$, i.e., they can be written as:
\begin{subequations}\label{equ:7}
\begin{align}
& Y_f = \{{\bf y}_f \in \mathbb{R}^{p K_f}: G {\bf y}_f \le g, H {\bf y}_f = h\} \\
& c(u(t), y(t)) = \begin{bmatrix} S_u & S_y \end{bmatrix} \begin{bmatrix} u(t) \\ y(t) \end{bmatrix} - s \le 0 
\end{align}
\end{subequations}
where $G, H, S_u, S_y$ and $g, h, s$ are matrices and vectors of compatible dimensions. Furthermore, we make the following ``controlled invariance'' assumption about $Y_f$:

{\it Assumption~3:} Let $\bar{K}_f = \max(K_f, l_{\min})$. For any admissible pair of input-output trajectories of length $\bar{K}_f$, $({\bf u}_{[0:\bar{K}_f-1]},{\bf y}_{[0:\bar{K}_f-1]})$, that satisfies the path constraints $c(u(t), y(t)) \le 0$ for $t = 0, ..., \bar{K}_f-1$ and the target condition ${\bf y}_{[\bar{K}_f-K_f:\bar{K}_f-1]} \in Y_f$, there exist an input $u(\bar{K}_f)$ and its corresponding output~$y(\bar{K}_f)$ such that they satisfy $c(u(\bar{K}_f), y(\bar{K}_f)) \le 0$ and ${\bf y}_{[\bar{K}_f-K_f+1:\bar{K}_f]} \in Y_f$.

In {\it Assumption~3}, because $\bar{K}_f \ge l_{\min}$, an admissible pair of input-output trajectories $({\bf u}_{[0:\bar{K}_f-1]},{\bf y}_{[0:\bar{K}_f-1]})$ corresponds to a unique initial state $x(0)$ and a unique state trajectory $x(0),x(1),...,x(\bar{K}_f)$. Hence, for an input $u(\bar{K}_f)$, the corresponding output $y(\bar{K}_f)$ is also unique. {\it Assumption~3} means that for any pair of input-state trajectories the output trajectory of which enters the target set $Y_f$ while satisfying the path constraints, there exists a control to maintain the output trajectory in $Y_f$ while satisfying the path constraints. Hence, it specifies a ``controlled invariance'' property of $Y_f$ with respect to the system dynamics and the path~constraints.

{\it Remark~1:} While in many conventional trajectory optimization problem formulations the initial condition is a~given value $x_i$ for the state at the initial time $t = 0$, a given value for the output or a given pair of input-output at a single time is not sufficient for uniquely determining the trajectory. Therefore, we consider a pair of input-output trajectories of length $K_i$ that satisfies $K_i \ge l_{\min}$ as the initial condition which uniquely determines the initial state according to \eqref{equ:6} and hence the trajectory. For the terminal condition, we consider a target set $Y_f$ instead of a single point to represent a larger class of problems which includes a single target point as a special case.

Lastly, we assume that a state-space model of the considered system (i.e., the matrices $(A, B, C, D)$ in \eqref{equ:1}) is not given and only input-output trajectory data of the system are available. For instance, we may deal with a real system that has uncertain parameters while can generate input-output data (see the example in~Section~\ref{sec:5}). The goal of this paper is to develop a computationally-efficient approach to the minimum-time trajectory optimization problem~\eqref{equ:5} for such a~setting. We note that although it is possible to first identify the matrices $(A, B, C, D)$ using input-output data and system identification techniques and then solve the problem~\eqref{equ:5} based on the identified state-space model, this two-step approach may be cumbersome from the point of view of practitioners and $(A, B, C, D)$ that is compatible with given data is in general not unique. Therefore, we pursue an end-to-end solution -- directly from input-output data to a solution to~\eqref{equ:5}. 

% Our approach is based on the fundamental lemma of \cite{willems2005note} and a continuous optimization-based scheme for time-optimal control proposed in \cite{verschueren2017stabilizing}.

\section{Data-based Model for Long-term Trajectory Prediction}\label{sec:3}

Assume that we have input-output trajectory data of length~$M$:
\begin{equation}\label{equ:8}
\mathcal{D}_M = \left\{\begin{bmatrix} u^d(0) \\ y^d(0) \end{bmatrix}, \begin{bmatrix} u^d(1) \\ y^d(1) \end{bmatrix}, ..., \begin{bmatrix} u^d(M-1) \\ y^d(M-1) \end{bmatrix}\right\}
\end{equation}
where the superscript $d$ indicates ``data.'' Note that these input-output pairs $(u^d(t),y^d(t))$, $t = 0,...,M-1$, should be sampled from a single trajectory\footnote{If the data are from multiple trajectories, then the inputs shall satisfy a {\it collectively persistently exciting} condition \citep{van2020willems}.}. Construct the following Hankel data matrices:
\begin{subequations}\label{equ:9}
\begin{align}
&\! \mathcal{H}_L(u^d) \!=\! \begin{bmatrix} u^d(0) & u^d(1) & \cdots & u^d(M-L) \\
u^d(1) & u^d(2) & \cdots & u^d(M-L+1) \\
\vdots & \vdots & \ddots & \vdots \\
u^d(L-1) & u^d(L) & \cdots & u^d(M-1) \end{bmatrix} \\
&\! \mathcal{H}_L(y^d) \!=\! \begin{bmatrix} y^d(0) & y^d(1) & \cdots & y^d(M-L) \\
y^d(1) & y^d(2) & \cdots & y^d(M-L+1) \\
\vdots & \vdots & \ddots & \vdots \\
y^d(L-1) & y^d(L) & \cdots & y^d(M-1) \end{bmatrix} \tag{9b}
\end{align}
\end{subequations}
where $L$ indicates the number of stacks of the signal $u^d$ or $y^d$ in each column. The control input trajectory $u^d(0), u^d(1), ..., u^d(M-1)$ is said to be {\it persistently
exciting of order $L$} if $\mathcal{H}_L(u^d)$ has full rank.

Our approach uses a result known as the fundamental lemma of \cite{willems2005note}. The following lemma is an equivalent statement of Willems' fundamental lemma in a state-space
context~\citep{van2020willems}:

{\it Lemma~1:} If the system \eqref{equ:1} is controllable and the control input trajectory $u^d(0), u^d(1), ..., u^d(M-1)$ is persistently exciting of order $L + n$, then a pair of input-output trajectories of length $L$, $({\bf u}_{[t:t+L-1]},{\bf y}_{[t:t+L-1]})$, is admissible by the system~\eqref{equ:1} if and only if
\begin{equation}\label{equ:10}
\begin{bmatrix} {\bf u}_{[t:t+L-1]} \\ {\bf y}_{[t:t+L-1]} \end{bmatrix} = \begin{bmatrix} \mathcal{H}_L(u^d) \\ \mathcal{H}_L(y^d) \end{bmatrix} \zeta
\end{equation}
for some vector $\zeta \in \mathbb{R}^{M-L+1}$.

Assume $L > K_i$ and let $t = -K_i$. Then, \eqref{equ:10} can be written as
\begin{equation}\label{equ:11}
\begin{bmatrix} {\bf u}_{[-K_i:-1]} \\ {\bf u}_{[0:L-K_i-1]} \\ {\bf y}_{[-K_i:-1]} \\ {\bf y}_{[0:L-K_i-1]} \end{bmatrix} = \begin{bmatrix} \mathcal{H}_L(u^d) \\ \mathcal{H}_L(y^d) \end{bmatrix} \zeta
\end{equation}
Given $({\bf u}_{[-K_i:-1]},{\bf y}_{[-K_i:-1]}) = ({\bf u}_i,{\bf y}_i)$, \eqref{equ:11} is a linear equation of variables $({\bf u}_{[0:L-K_i-1]},{\bf y}_{[0:L-K_i-1]}, \zeta) \in \mathbb{R}^{m(L-K_i)} \times \mathbb{R}^{p(L-K_i)} \times \mathbb{R}^{M-L+1}$. In \cite{coulson2019data}, \eqref{equ:11} is used as a model for predicting the outputs $y(t)$ over the entire planning horizon $t = 0, ..., N-1$. This requires $L = N + K_i$. When the planning horizon~$N$ is large, which is not rare in a trajectory optimization setting, such a strategy requires high-dimensional data matrices $(\mathcal{H}_L(u^d),\mathcal{H}_L(y^d))$ and requires the control input trajectory $u^d(0), u^d(1), ..., u^d(M-1)$ to be persistently exciting of an order of $L + n = N + K_i + n$. Inspired by the multiple shooting method for trajectory optimization over an extended planning horizon~\citep{trelat2012optimal}, we consider partitioning the entire trajectory of length $N = K(L-K_i)$ into $K$ segments:
\begin{subequations}\label{equ:12}
\begin{align} 
& {\bf u}_{[0:N-1]} = \begin{bmatrix} {\bf u}_{[0:L-K_i-1]} \\ {\bf u}_{[L-K_i:2(L-K_i)-1]} \\ \vdots \\ {\bf u}_{[(K-1)(L-K_i):N-1]} \end{bmatrix} \\
& {\bf y}_{[0:N-1]} = \begin{bmatrix} {\bf y}_{[0:L-K_i-1]} \\ {\bf y}_{[L-K_i:2(L-K_i)-1]} \\ \vdots \\ {\bf y}_{[(K-1)(L-K_i):N-1]} \end{bmatrix}
\end{align}
\end{subequations}
For each segment $k$, $k = 1,...,K$, we stack its previous $K_i$ points on top to get the following vectors:
\begin{subequations}\label{equ:13}
\begin{align}
{\sf u}_k &= \begin{bmatrix} {\bf u}_{[(k-1)(L-K_i)-K_i:(k-1)(L-K_i)-1]} \\ {\bf u}_{[(k-1)(L-K_i):k(L-K_i)-1]} \end{bmatrix} \in \mathbb{R}^{mL} \\ 
{\sf y}_k &= \begin{bmatrix} {\bf y}_{[(k-1)(L-K_i)-K_i:(k-1)(L-K_i)-1]} \\ {\bf y}_{[(k-1)(L-K_i):k(L-K_i)-1]} \end{bmatrix} \in \mathbb{R}^{pL}
\end{align}
\end{subequations}
According to {\it Lemma~1}, for each $k = 1,...,K$, the pair $({\sf u}_k, {\sf y}_k)$ is an admissible pair of input-output trajectories if and only if
\begin{equation}\label{equ:14}
\begin{bmatrix} {\sf u}_k \\ {\sf y}_k \end{bmatrix} = \begin{bmatrix} \mathcal{H}_L(u^d) \\ \mathcal{H}_L(y^d) \end{bmatrix} \zeta_k
\end{equation}
for some $\zeta_k \in \mathbb{R}^{M-L+1}$. We refer to \eqref{equ:14} for $k = 1,...,K$ as equality {\it dynamic constraints} with $\zeta_k$ as auxiliary variables. Then, we impose the following equality {\it matching conditions} for $k = 1,...,K-1$ to piece together the segments and form a long admissible trajectory:
\begin{subequations}\label{equ:15}
\begin{align}
\begin{bmatrix} I_{m K_i} \\ 0_{m(L-K_i),m K_i} \end{bmatrix}^{\top} {\sf u}_{k+1} &= \begin{bmatrix} 0_{m(L-K_i),m K_i} \\ I_{m K_i} \end{bmatrix}^{\top} {\sf u}_k \\
\begin{bmatrix} I_{p K_i} \\ 0_{p(L-K_i),p K_i} \end{bmatrix}^{\top} {\sf y}_{k+1} &= \begin{bmatrix} 0_{p(L-K_i),p K_i} \\ I_{p K_i} \end{bmatrix}^{\top} {\sf y}_k
\end{align}
\end{subequations}
i.e., we enforce the first $K_i$ points of ${\sf u}_{k+1}$ (resp. ${\sf y}_{k+1}$) to be equal to the last $K_i$ points of ${\sf u}_k$ (resp. ${\sf y}_k$). Note that because $K_i \ge l_{\min}$, an admissible pair of input-output trajectories of length $K_i$ corresponds to a unique state trajectory of length $K_i + 1$. Specifically, given a state-space model \eqref{equ:1} of the system, for an admissible pair of input-output trajectories of length $K_i$, with $K_i \ge l_{\min}$, the unique corresponding initial state can be determined by \eqref{equ:4} with $l = K_i$, and then the states over the next $K_i$ steps are uniquely determined by this initial state and the given input trajectory. Therefore, matching the first $K_i$ points of $({\sf u}_{k+1},{\sf y}_{k+1})$ to the last $K_i$ points of $({\sf u}_k,{\sf y}_k)$ creates a continuous state trajectory $x((k-1)(L-K_i)-K_i), ..., x((k+1)(L-K_i))$.

\begin{table*}[bp]
\centering
\begin{minipage}{1\textwidth}
\hrule
\begin{subequations}\label{equ:18}
\begin{align}
{\bf u}_{[-K_i:N-1]} &= \text{col}\left(\begin{bmatrix} I_{m K_i} \\ 0_{m(L-K_i),m K_i} \end{bmatrix}^{\top} {\sf u}_1, \begin{bmatrix} 0_{m K_i,m(L-K_i)} \\ I_{m(L-K_i)} \end{bmatrix}^{\top} {\sf u}_1, ..., \begin{bmatrix} 0_{m K_i,m(L-K_i)} \\ I_{m(L-K_i)} \end{bmatrix}^{\top} {\sf u}_K \right) \tag{18a} \\ 
{\bf y}_{[-K_i:N-1]} &= \text{col}\left(\begin{bmatrix} I_{p K_i} \\ 0_{p(L-K_i),p K_i} \end{bmatrix}^{\top} {\sf y}_1, \begin{bmatrix} 0_{p K_i, p(L-K_i)} \\ I_{p(L-K_i)} \end{bmatrix}^{\top} {\sf y}_1, ..., \begin{bmatrix} 0_{p K_i,p(L-K_i)} \\ I_{p(L-K_i)} \end{bmatrix}^{\top} {\sf y}_K \right) \tag{18b}
\end{align}
\end{subequations}
\hrule
\end{minipage}
\end{table*}
\setcounter{equation}{15}

\begin{figure}[htbp!]
\begin{center}
\begin{picture}(240.0, 112.5)
\put(-5,0){\epsfig{file=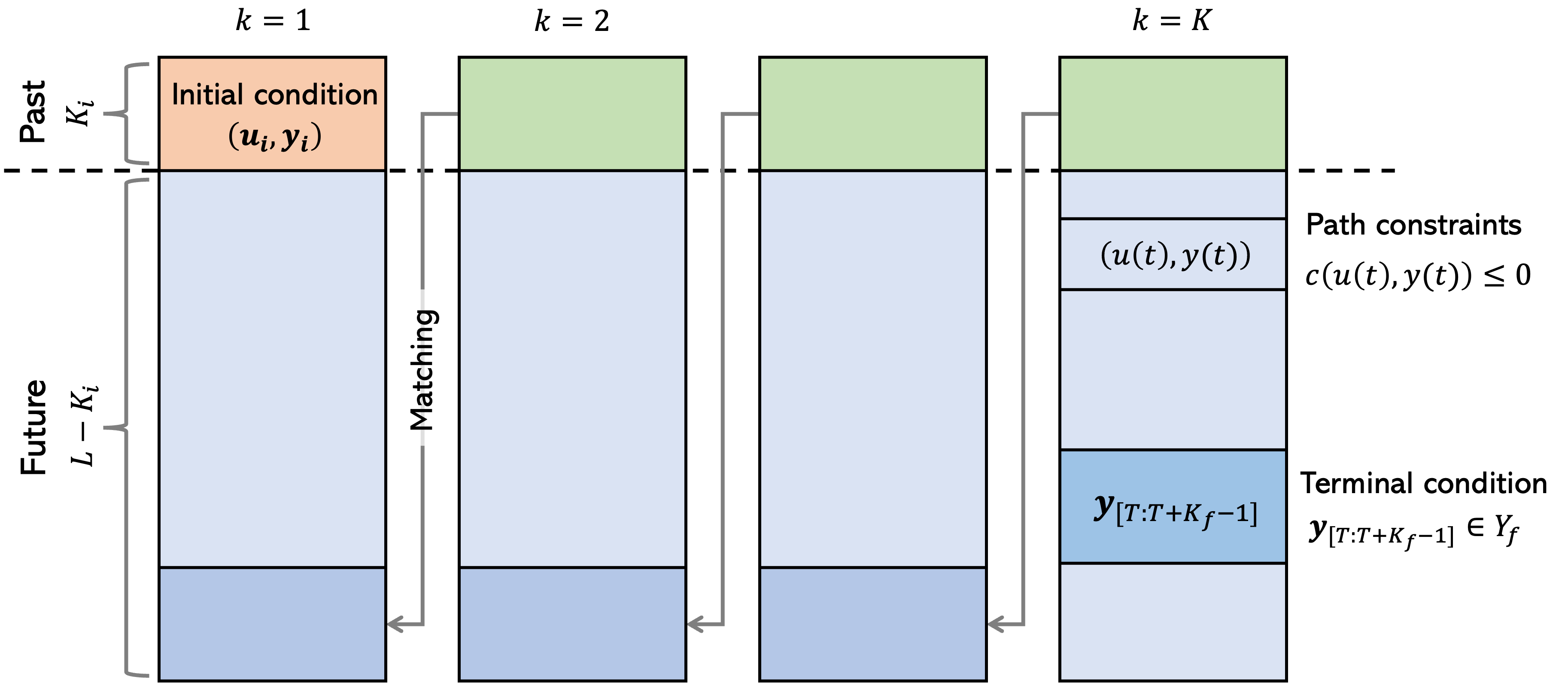,height=.225\textwidth}}
\end{picture}
\end{center}
      \caption{Illustration of the structure of the variables.}
      \label{fig:1}
\end{figure}

The initial condition for the input-output trajectories in \eqref{equ:52} can be imposed through the following equality constraints on $({\sf u}_1, {\sf y}_1)$:
\begin{subequations}\label{equ:16}
\begin{align}
\begin{bmatrix} I_{m K_i} \\ 0_{m(L-K_i),m K_i} \end{bmatrix}^{\top} {\sf u}_1 &= {\bf u}_i \\
\begin{bmatrix} I_{p K_i} \\ 0_{p(L-K_i),p K_i} \end{bmatrix}^{\top} {\sf y}_1 &= {\bf y}_i
\end{align}
\end{subequations}
Also, the path constraints in \eqref{equ:54} can be imposed through the following inequality constraints for $k = 1,...,K$ and $l = K_i+1, ..., L$:
\begin{equation}\label{equ:17}
S_u \begin{bmatrix} 0_{m (l-1),m} \\ I_m \\ 0_{m(L-l),m} \end{bmatrix}^{\top} {\sf u}_k + S_y \begin{bmatrix} 0_{p (l-1),p} \\ I_p \\ 0_{p(L-l),p} \end{bmatrix}^{\top} {\sf y}_k \le s 
\end{equation}
\setcounter{equation}{18}

The input-output trajectories over the entire planning horizon $t = -K_i, ..., N-1$ can be constructed from the vectors ${\sf u}_k$ and~${\sf y}_k$, $k = 1,...,K$, through the equations in (18). And we arrive at the following result:

{\it Lemma~2:} Suppose the system \eqref{equ:1} is controllable and the control input trajectory $u^d(0), u^d(1), ..., u^d(M-1)$ is persistently exciting of order $L + n$. Then, a pair of input-output trajectories $({\bf u}_{[-K_i:N-1]},{\bf y}_{[-K_i:N-1]})$, where $N$ can be written as $N = K(L-K_i)$ for some $K \in \mathbb{Z}_{\ge 1}$, is admissible by the system~\eqref{equ:1} and satisfies the initial condition in \eqref{equ:52} and the path constraints in \eqref{equ:54} if and only if there exist vectors $({\sf u}_k, {\sf y}_k, \zeta_k) \in \mathbb{R}^{mL} \times \mathbb{R}^{pL} \times \mathbb{R}^{M-L+1}$, $k = 1,...,K$, that satisfy the constraints in~\eqref{equ:14}, \eqref{equ:15}, \eqref{equ:16} and \eqref{equ:17} and $({\bf u}_{[-K_i:N-1]},{\bf y}_{[-K_i:N-1]})$ relate to the vectors $({\sf u}_k, {\sf y}_k)$, $k = 1,...,K$, according to (18).

{\it Proof:} This result follows from {\it Lemma~1} and the constructions of the vectors $({\sf u}_k, {\sf y}_k)$, $k = 1,...,K$, in \eqref{equ:12}--\eqref{equ:13} and the constraints in \eqref{equ:14}--\eqref{equ:17}. $\blacksquare$

\section{Linear Program for Minimum-time Trajectory Optimization}\label{sec:4}

Now we deal with the target condition ${\bf y}_{[T:T+K_f-1]} \in Y_f$. Recall that the goal is to minimize the time $T$ at which the output trajectory reaches the target set $Y_f = \{{\bf y}_f \in \mathbb{R}^{p K_f}: G {\bf y}_f \le g, H {\bf y}_f = h\}$.

Assume that the minimum time $T^*$ is in the range of $[T_0, T_1]$. For a given problem, such a range may be estimated based on prior knowledge or set to be sufficiently large (e.g., $T_0 = 0$ and $T_1$ being a large number). It will become clear soon that a smaller range (i.e., a closer estimate of $T^*$) makes the formulated problem simpler in terms of having less decision variables and constraints. For each $t \in [T_0, T_1]$, define a slack variable $\varepsilon_t \in \mathbb{R}^{q_g + q_h}$, where $q_g$ is the dimension of $g$ and $q_h$ is that of $h$. Impose the following constraints for $\varepsilon_t$:
\begin{subequations}\label{equ:19}
\begin{align}
& G {\bf y}_{[t:t+K_f-1]} - g \le \varepsilon_{t,1:q_g} \\
& H {\bf y}_{[t:t+K_f-1]} - h = \varepsilon_{t,q_g+1:q_g+q_h}
\end{align}
\end{subequations}
where $\varepsilon_{t,1:q_g}$ denotes the first $q_g$ rows of $\varepsilon_t$ and $\varepsilon_{t,q_g+1:q_g+q_h}$ denotes the remaining rows of $\varepsilon_t$. Then, ${\bf y}_{[t:t+K_f-1]} \in Y_f$ if and only if $0$ is a feasible value for~$\varepsilon_t$. 

Consider the following function:
\begin{equation}\label{equ:20}
J = \sum_{t = T_0}^{T_1} \theta^{t-T_0} \|\varepsilon_t\|_1
\end{equation}
where $\theta > 1$ is a sufficiently large constant and $\|\cdot\|_1$ denotes the $1$-norm. The analysis in what follows shows that minimizing \eqref{equ:20} can lead to a minimum-time trajectory solution: 

Assume $T^* \in [T_0, T_1]$ is known and consider the following problem parameterized by $\eta = \{\eta_t\}_{t = T^*, ..., T_1}$:
\begin{subequations}\label{equ:21}
\begin{align}
\min_{u(\cdot), y(\cdot), \varepsilon_{(\cdot)}}\!\! &\, J_0(\theta,\{\varepsilon_t\}_{t = T_0,...,T^*-1}) = \sum_{t = T_0}^{T^*-1} \theta^{t-T_0} \|\varepsilon_t\|_1 \label{equ:211} \\
\text{s.t.}\quad &\, ({\bf u}_{[-K_i:-1]},{\bf y}_{[-K_i:-1]}) = ({\bf u}_i,{\bf y}_i) \label{equ:212} \\
&\, c(u(t), y(t)) \le 0, \,\,\, t = 0, ..., N-1 \label{equ:213} \\
&\, G {\bf y}_{[t:t+K_f-1]} - g \le \varepsilon_{t,1:q_g} \label{equ:214} \\
&\, H {\bf y}_{[t:t+K_f-1]} - h = \varepsilon_{t,q_g+1:q_g+q_h}, \,\, t = T_0, ..., T_1 \label{equ:215} \\
&\, \varepsilon_t = \eta_t, \,\,\, t = T^*, ..., T_1 \label{equ:216}
\end{align}
\end{subequations}
where $N$ represents a planning horizon that satisfies $N \ge T_1 + K_f$, and $\eta_t \in \mathbb{R}^{q_g + q_h}$, $t = T^*, ..., T_1$, are parameters with the nominal value $\eta_t^* = 0$. The following result clarifies the relation between the minimum-time problem of interest, \eqref{equ:5}, and the above problem \eqref{equ:21}:

{\it Theorem~1:} (i) Suppose {\it Assumptions~2} and {\it 3} hold. Let $({\bf u}_{[-K_i:T^*+K_f-1]}^*,{\bf y}_{[-K_i:T^*+K_f-1]}^*,T^*)$ be an optimal solution to the minimum-time problem \eqref{equ:5} with $T^* \in [T_0, T_1]$. Then, \eqref{equ:21} with $\eta = 0$ has a feasible solution $({\bf u}_{[-K_i:N-1]}',{\bf y}_{[-K_i:N-1]}',\{\varepsilon_t'\}_{t = T_0,...,T_1})$ that satisfies $({\bf u}_{[-K_i:T^*+K_f-1]}',{\bf y}_{[-K_i:T^*+K_f-1]}') = ({\bf u}_{[-K_i:T^*+K_f-1]}^*,{\bf y}_{[-K_i:T^*+K_f-1]}^*)$ and $\varepsilon_t' = 0$ for $t = T^*, ...,T_1$.

(ii) Suppose \eqref{equ:5} is feasible and has a minimum time $T^* \in [T_0, T_1]$. Let $({\bf u}_{[-K_i:N-1]}',{\bf y}_{[-K_i:N-1]}',\{\varepsilon_t'\}_{t = T_0,...,T_1})$ be an arbitrary feasible solution to \eqref{equ:21} with $\eta = 0$. Then, the triple $({\bf u}_{[-K_i:T^*+K_f-1]}',{\bf y}_{[-K_i:T^*+K_f-1]}',T^*)$ is an optimal solution to \eqref{equ:5}.

{\it Proof:} For (i), let $({\bf u}_{[-K_i:T^*+K_f-1]}^*,{\bf y}_{[-K_i:T^*+K_f-1]}^*,T^*)$ be an optimal solution to \eqref{equ:5}. Because $K_i \ge l_{\min}$ and $T^* \ge 0$, $\bar{K}_f = \max(K_f, l_{\min})$ must satisfy $-K_i \le T^*+K_f-\bar{K}_f \le T^*+K_f-1$. Then, under {\it Assumption~2}, $({\bf u}_{[T^*+K_f-\bar{K}_f:T^*+K_f-1]}^*,{\bf y}_{[T^*+K_f-\bar{K}_f:T^*+K_f-1]}^*)$ is an admissible pair of input-output trajectories of length $\bar{K}_f$ that satisfies $c(u(t), y(t)) \le 0$ for $t = T^*+K_f-\bar{K}_f, ..., T^*+K_f-1$ and ${\bf y}_{[T^*:T^*+K_f-1]} \in Y_f$. In this case, under {\it Assumption~3}, there exist inputs $u'(T^*+K_f), ..., u'(N-1)$ and the corresponding outputs $y'(T^*+K_f), ..., y'(N-1)$ such that $c(u(t), y(t)) \le 0$ for $t = T^*+K_f, ..., N-1$ and ${\bf y}_{[t:t+K_f-1]} \in Y_f$ for $t = T^* + 1, ..., T_1$. Now let ${\bf u}_{[-K_i:N-1]}' = \text{col}({\bf u}_{[-K_i:T^*+K_f-1]}^*,u'(T^*+K_f),...,u'(N-1))$ and ${\bf y}_{[-K_i:N-1]}' = \text{col}({\bf y}_{[-K_i:T^*+K_f-1]}^*,y'(T^*+K_f),...,y'(N-1))$, and let $\varepsilon_t'$ be defined according to $\varepsilon_{t,1:q_g}' = G {\bf y}_{[t:t+K_f-1]}' - g$ and $\varepsilon_{t,q_g+1:q_g+q_h}' = H {\bf y}_{[t:t+K_f-1]}' - h$ for $t = T_0, ..., T^*-1$ and $\varepsilon_t' = 0$ for $t = T^*, ...,T_1$. Then, the triple $({\bf u}_{[-K_i:N-1]}',{\bf y}_{[-K_i:N-1]}',\{\varepsilon_t'\}_{t = T_0,...,T_1})$ is a feasible solution to \eqref{equ:21} with $\eta = 0$. This proves part (i).

For part (ii), let $({\bf u}_{[-K_i:N-1]}',{\bf y}_{[-K_i:N-1]}',\{\varepsilon_t'\}_{t = T_0,...,T_1})$ be a feasible solution to \eqref{equ:21} with $\eta = 0$. Due to the constraints in \eqref{equ:214}--\eqref{equ:216}, this solution satisfies ${\bf y}_{[T^*:T^*+K_f-1]}' \in Y_f$. Therefore, the triple $({\bf u}_{[-K_i:T^*+K_f-1]}',{\bf y}_{[-K_i:T^*+K_f-1]}',T^*)$ is an optimal solution to the minimum-time problem \eqref{equ:5}. $\blacksquare$

{\it Remark~2:} From the proof of {\it Theorem~1} the necessity of both {\it Assumptions~2} and {\it 3} for guaranteeing the result of part (i) should be clear. For instance, suppose the second half of {\it Assumption~2} does not hold, i.e., the pair $({\bf u}_i,{\bf y}_i)$ does not satisfy the path constraints $c(u(t), y(t)) \le 0$ at all times, and suppose $K_f < l_{\min}$ and $T^* < l_{\min} - K_f$. Then, the pair $({\bf u}_{[T^*+K_f-\bar{K}_f:T^*+K_f-1]}^*,{\bf y}_{[T^*+K_f-\bar{K}_f:T^*+K_f-1]}^*)$ may not satisfy $c(u(t), y(t)) \le 0$ for all $t = T^*+K_f-\bar{K}_f, ..., T^*+K_f-1$. Consequently, for such a pair of input-output trajectories of length $\bar{K}_f$, there may not exist an input $u'(T^*+K_f)$ that is able to maintain the output trajectory in $Y_f$ while satisfying the path constraints even if {\it Assumption~3} holds. In contrast, the result of part (ii) does not rely on {\it Assumptions~2} and {\it 3}.

{\it Theorem~1} indicates that minimum-time trajectory solutions (i.e., optimal solutions to \eqref{equ:5}) can be obtained through~\eqref{equ:21}. However, the formulation of \eqref{equ:21} relies on the exact knowledge of the minimum time $T^*$, which is typically not known a priori (otherwise the problem reduces to a fixed-time trajectory planning problem). In what follows we show that an optimal solution to~\eqref{equ:21} can be obtained through another related problem the formulation of which does not rely on exact knowledge of~$T^*$. The technique originates from exact penalty methods for handling constraints in constrained optimization.

Now let $({\bf u}_{[-K_i:N-1]}^*(\theta),{\bf y}_{[-K_i:N-1]}^*(\theta),\{\varepsilon_t^*(\theta)\}_{t = T_0,...,T_1})$ be an optimal solution to \eqref{equ:21} with $\eta = \{\eta_t\}_{t = T^*, ..., T_1} = 0$ and a certain value of $\theta$. Let $\lambda_k(\theta) \in \mathbb{R}^{q_g + q_h}$ be the Lagrange multiplier associated with the constraint $\varepsilon_k = \eta_k$, $k = T^*, ..., T_1$. Its value satisfies \begin{equation}\label{equ:22}
\lambda_{k,i}(\theta) \!=\! \frac{\text{d}J_0(\theta,\{\varepsilon_t^*(\theta,\eta_{k,i})\}_{t = T_0,...,T^*-1})}{\text{d}\eta_{k,i}},\,\, i \!=\! 1,...,q_g + q_h
\end{equation}
where $\varepsilon_t^*(\theta,\eta_{k,i})$ denotes the perturbed value of $\varepsilon_t^*(\theta)$ due to a perturbation $\eta_{k,i} \in \mathbb{R}$ to the $i$th entry of the parameter $\eta_k$, i.e., the Lagrange multiplier represents the sensitivity of the cost to perturbations to the constraint \citep{buskens2001sensitivity}. We~make~the following assumption about the optimal solution $({\bf u}_{[-K_i:N-1]}^*(\theta),{\bf y}_{[-K_i:N-1]}^*(\theta),\{\varepsilon_t^*(\theta)\}_{t = T_0,...,T_1})$ to~\eqref{equ:21}:

{\it Assumption~4:} For sufficiently large $\theta$ and $t = T_0,...,T^*-1$, the sensitivity of $\varepsilon_t^*(\theta)$ to perturbations to the constraints $\varepsilon_k = \eta_k$, $k = T^*, ..., T_1$, is bounded by a constant~$R$, i.e.,
\begin{equation}\label{equ:23}
\left\|\frac{\text{d}\varepsilon_t^*(\theta,\eta_{k,i})}{\text{d}\eta_{k,i}}\right\|_1 \le R, \quad i = 1,...,q_g + q_h
\end{equation}
where $\|\cdot\|_1$ denotes the $1$-norm.

Under {\it Assumption~4}, we can derive the following bound on $\lambda_{k,i}(\theta)$:
\begin{align}\label{equ:24}
& |\lambda_{k,i}(\theta)| = \left|\frac{\text{d}J_0(\theta,\{\varepsilon_t^*(\theta,\eta_{k,i})\}_{t = T_0,...,T^*-1})}{\text{d}\eta_{k,i}}\right| \nonumber \\
&\le \sum_{t = T_0}^{T^*-1} \left|\frac{\partial J_0}{\partial \varepsilon_t^*} \cdot \frac{\partial \varepsilon_t^*}{\partial \eta_{k,i}}\right| \le \sum_{t = T_0}^{T^*-1} \theta^{t-T_0} \left\|\frac{\text{d}\varepsilon_t^*}{\text{d}\eta_{k,i}}\right\|_1 \nonumber \\
&\le R \sum_{t = T_0}^{T^*-1} \theta^{t-T_0} \le \frac{R}{\theta-1} \theta^{T^*-T_0}
\end{align}
Hence, if $\theta \ge R + 1$, we have
\begin{equation}\label{equ:25}
\|\lambda_k(\theta)\|_{\infty} = \max_i |\lambda_{k,i}(\theta)| \le \theta^{T^*-T_0}
\end{equation}
for $k = T^*, ..., T_1$, where $\|\cdot\|_{\infty}$ denotes the sup-norm. We arrive at the following result:

{\it Theorem~2:} Suppose \eqref{equ:5} is feasible and has a minimum time $T^* \in [T_0, T_1]$. Let $({\bf u}_{[-K_i:N-1]}^*(\theta),{\bf y}_{[-K_i:N-1]}^*(\theta),$ $\{\varepsilon_t^*(\theta)\}_{t = T_0,...,T_1})$ be an optimal solution to \eqref{equ:21} with $\eta~=~0$ and $\theta > 1$. Suppose {\it Assumption~4} holds and $\theta$ is sufficiently large. Then, $({\bf u}_{[-K_i:N-1]}^*(\theta),{\bf y}_{[-K_i:N-1]}^*(\theta),$ $\{\varepsilon_t^*(\theta)\}_{t = T_0,...,T_1})$ is an optimal solution to the following problem, the formulation of which does not rely on $T^*$:
\begin{subequations}\label{equ:26}
\begin{align}
\min_{u(\cdot), y(\cdot), \varepsilon_{(\cdot)}}\!\! &\, J = \sum_{t = T_0}^{T_1} \theta^{t-T_0} \|\varepsilon_t\|_1 \label{equ:261} \\
\text{s.t.}\quad &\, ({\bf u}_{[-K_i:-1]},{\bf y}_{[-K_i:-1]}) = ({\bf u}_i,{\bf y}_i) \label{equ:262} \\
&\, c(u(t), y(t)) \le 0, \,\,\, t = 0, ..., N-1 \label{equ:263} \\
&\, G {\bf y}_{[t:t+K_f-1]} - g \le \varepsilon_{t,1:q_g} \label{equ:264} \\
&\, H {\bf y}_{[t:t+K_f-1]} - h = \varepsilon_{t,q_g+1:q_g+q_h}, \,\, t = T_0, ..., T_1 \label{equ:265}
\end{align}
\end{subequations}

{\it Proof:} The cost function of \eqref{equ:26} can be written as $J = J_0 + \sum_{t = T^*}^{T_1} \theta^{t-T_0} \|\varepsilon_t\|_1$, where $J_0$ is the cost function of~\eqref{equ:21}. If {\it Assumption~4} holds and $\theta \ge R + 1$, then according to \eqref{equ:25}, for $t = T^*, ..., T_1$, the Lagrange multiplier $\lambda_t(\theta)$ associated with the constraint $\varepsilon_t = \eta_t = 0$ of~\eqref{equ:21} satisfies $\|\lambda_t(\theta)\|_{\infty} \le \theta^{T^*-T_0} \le \theta^{t-T_0}$. This implies that the term $\theta^{t-T_0} \|\varepsilon_t\|_1$ in the cost function of~\eqref{equ:26} is an exact penalty for the constraint $\varepsilon_t = 0$. Therefore, \eqref{equ:26} is a reformulation of \eqref{equ:21} by replacing all constraints in \eqref{equ:216} with exact penalties and hence the result follows \citep{han1979exact}. $\blacksquare$

{\it Theorem~2} indicates that optimal solutions to \eqref{equ:21}, which, according to {\it Theorem~1}, are optimal solutions to~\eqref{equ:5} (i.e., minimum-time trajectories), can be obtained through~\eqref{equ:26}. The cost function of \eqref{equ:26}, which is non-smooth due to the $1$-norms, can be readily converted into a smooth function using a linear programming reformulation. We combine the results of the previous section on data-based trajectory prediction and of this section on minimum-time trajectory planning and arrive at the following problem formulation:
\begin{align}\label{equ:27}
& \min_{\{{\sf u}_k, {\sf y}_k, \zeta_k\}_{k=1,...K}, \{\varepsilon_t \ge 0\}_{t=T_0,...,T_1}}\! J = \sum_{t = T_0}^{T_1} \theta^{t-T_0} (1_{q_g+q_h}^{\top} \varepsilon_t) \nonumber \\
&\quad\quad\quad\quad\quad\,\,\, \text{s.t.} \\
& \text{dynamic constraints: \eqref{equ:14} for } k = 1,...,K \nonumber \\
& \text{matching conditions: \eqref{equ:15} for } k = 1,...,K-1 \nonumber \\
& \text{initial condition: \eqref{equ:16}} \nonumber \\
& \text{path constraints: \eqref{equ:17} for } k = 1,..,K \text{ and } l = K_i+1, ..., L \nonumber \\
& \text{terminal condition: } \nonumber \\
& G {\bf y}_{[t:t+K_f-1]} - g \le \varepsilon_{t,1:q_g} \nonumber \\
& H {\bf y}_{[t:t+K_f-1]} - h \le \varepsilon_{t,q_g+1:q_g+q_h} \nonumber \\
& h - H {\bf y}_{[t:t+K_f-1]} \le \varepsilon_{t,q_g+1:q_g+q_h} \text{ for } t = T_0, ..., T_1 \nonumber
\end{align}
in which the dimensions of the decision variables are ${\sf u}_k \in \mathbb{R}^{mL}$, ${\sf y}_k \in \mathbb{R}^{pL}$, $\zeta_k \in \mathbb{R}^{M-L+1}$, $\varepsilon_t \in \mathbb{R}^{q_g + q_h}$, and the number $K$ should be chosen as $K = \lceil \frac{T_1+K_f}{L-K_i} \rceil$, where $\lceil \cdot \rceil$ means rounding up to the nearest integer. As discussed in the second paragraph of this section, the range $[T_0,T_1]$ may be estimated based on prior knowledge about the problem or set as $[T_0,T_1] = [0,T']$ with $T'$ sufficiently large. It can be seen that the cost function of \eqref{equ:27} is a linear equation of decision variables and all constraints of \eqref{equ:27} are either linear equalities or linear inequalities of decision variables. Hence, \eqref{equ:27} is a linear program (LP) and can be easily solved using off-the-shelf LP solvers.

{\it Remark~3:} In \eqref{equ:27}, the auxiliary variables $\zeta_k \in \mathbb{R}^{M-L+1}$ only appear in the dynamic constraints \eqref{equ:14}. Using the following approach we can drop these variables from the problem: Let ${\sf v}_k = \text{col}({\sf u}_k, {\sf y}_k) \in \mathbb{R}^{(m+p)L}$ and $\mathcal{H} = \text{col}(\mathcal{H}_L(u^d),\mathcal{H}_L(y^d)) \in \mathbb{R}^{(m+p)L \times (M-L+1)}$. Then, \eqref{equ:14} can be written as ${\sf v}_k = \mathcal{H} \zeta_k$. Assume $\text{rank}(\mathcal{H}) = r$. We partition the rows of $\mathcal{H}$ into two groups: $\mathcal{H}_1 \in \mathbb{R}^{r \times (M-L+1)}$ collects $r$ linearly independent rows and $\mathcal{H}_2 \in \mathbb{R}^{((m+p)L-r) \times (M-L+1)}$ collects the remaining rows. Since $\text{rank}(\mathcal{H}) = r$, the rows of $\mathcal{H}_2$ are linear combinations of the rows of $\mathcal{H}_1$, i.e., $\mathcal{H}_2$ can be written as $\mathcal{H}_2 = \Gamma \mathcal{H}_1$, where $\Gamma \in \mathbb{R}^{((m+p)L-r) \times r}$ is uniquely determined by $\Gamma = \mathcal{H}_2 \mathcal{H}_1^\dagger$. Then, \eqref{equ:14} can be written as
\begin{equation}\label{equ:28}
\begin{bmatrix} {\sf v}_{k,1} \\ {\sf v}_{k,2} \end{bmatrix} = \begin{bmatrix} \mathcal{H}_1 \\ \mathcal{H}_2 \end{bmatrix} \zeta_k = \begin{bmatrix} \mathcal{H}_1 \zeta_k \\ \Gamma \mathcal{H}_1 \zeta_k \end{bmatrix} = \begin{bmatrix} \mathcal{H}_1 \zeta_k \\ \Gamma {\sf v}_{k,1} \end{bmatrix}
\end{equation}
where ${\sf v}_{k,1}$ (resp. ${\sf v}_{k,2}$) collects the entries of ${\sf v}_k$ corresponding to the rows of $\mathcal{H}_1$ (resp. $\mathcal{H}_2$), and the last equality is obtained by substituting ${\sf v}_{k,1} = \mathcal{H}_1 \zeta_k$ in the first row into the second row. Recall that, according to {\it Lemma~1}, ${\sf v}_k = \text{col}({\sf u}_k, {\sf y}_k)$ represents an admissible pair of input-output trajectories if and only if there exists $\zeta_k \in \mathbb{R}^{M-L+1}$ such that \eqref{equ:14} (equivalently, \eqref{equ:28}) holds. Since ${\sf v}_{k,1} \in \mathbb{R}^r$ and $\mathcal{H}_1$ has a rank of $r$, for any ${\sf v}_{k,1}$ there exists $\zeta_k$ such that the equation ${\sf v}_{k,1} = \mathcal{H}_1 \zeta_k$ in the first row of \eqref{equ:28} is satisfied. In this case, there exists $\zeta_k \in \mathbb{R}^{M-L+1}$ such that \eqref{equ:28} holds if and only if ${\sf v}_{k,1}$~and ${\sf v}_{k,2}$ satisfy the equation ${\sf v}_{k,2} = \Gamma {\sf v}_{k,1}$ in the second row of \eqref{equ:28}. Therefore, we can replace the dynamic constraints \eqref{equ:14} in \eqref{equ:27} with the linear-equality constraints ${\sf v}_{k,2} = \Gamma {\sf v}_{k,1}$ for $k = 1,...,K$, after which we obtain a linear program that is equivalent to \eqref{equ:27} while does not involve auxiliary variables $\zeta_k$. Note that the partition of $\mathcal{H} = \text{col}(\mathcal{H}_L(u^d),\mathcal{H}_L(y^d))$ into $\mathcal{H}_1$ and $\mathcal{H}_2$ and the calculation of $\Gamma = \mathcal{H}_2 \mathcal{H}_1^\dagger$ can be done offline and they are independent of $k$.

\section{Example: Spacecraft Relative Motion Planning}\label{sec:5}

To illustrate the approach, we consider a motion planning problem for a low-thrust spacecraft relative to a target body on a nominal circular orbit. The continuous-time dynamics are represented by the Clohessy-Wiltshire-Hill~(CWH) equations:
\begin{equation}\label{equ:29}
\dot{x} = A_c x + B_c u
\end{equation}
with
\begin{equation}\label{equ:30}
A_c = \begin{bmatrix} 0 & 0 & 0 & 1 & 0 & 0 \\
 0 & 0 & 0 & 0 & 1 & 0 \\
 0 & 0 & 0 & 0 & 0 & 1 \\
 3 \omega^2 & 0 & 0 & 0 & 2\omega & 0 \\
0 & 0 & 0 & -2\omega & 0 & 0 \\
0 & 0 & -\omega^2 & 0 & 0 & 0 
\end{bmatrix} \quad B_c = \begin{bmatrix} 0_{3,3} \\ \frac{T_{\max}}{m_s} I_3 
\end{bmatrix}
\end{equation}
where $\omega = \sqrt{\mu/r_o^3}$ is the orbital rate of the target (in radians/s), $\mu = 398,600$ km$^3$/s$^2$ is the gravitational parameter, $r_o = 6,928$ km is the radius of the target's circular orbit, $m_s = 50$ kg is the mass of the ego spacecraft, and $T_{\max} = 2 \times 10^{-4}$ kN represents the ego spacecraft's maximum thrust. The first three entries of the state vector $x \in \mathbb{R}^6$ represent the relative positions of the ego spacecraft with respect to the target body along the ${\sf x}$-, ${\sf y}$-, and ${\sf z}$-axes of the CWH frame (in~km), and the last three entries represent the relative velocity components (in km/s). The entries of the control input vector~$u \in \mathbb{R}^3$ represent the thrust level components of the ego spacecraft along the ${\sf x}$-, ${\sf y}$-, and ${\sf z}$-axes. For simplicity, we discretize the continuous-time model~\eqref{equ:29} using the forward Euler method with a sampling period of $\Delta t = 10$~s and obtain the discrete-time model:
\begin{equation}\label{equ:31}
x(t+1) = \underbrace{(I_6 + A_c \Delta t)}_A x(t) + \underbrace{(B_c \Delta t)}_B u(t)
\end{equation}
We treat the discrete-time model \eqref{equ:31} as the actual system and use it for both data generation and simulation. Also for simplicity, we assume a sup-norm bound on the control input vector: $\|u(t)\|_{\infty} \le 1$, which can be equivalently expressed as individual bounds on each thrust level component:
\begin{equation}\label{equ:32}
-1 \le u_i(t) \le 1, \quad i \in \{x,y,z\} 
\end{equation}
We remark that higher-fidelity modeling is possible: For piecewise-constant or piecewise-linear control inputs, exact discrete-time models can be obtained through the zero- or first-order hold method. For a 2-norm bound on the thrust level vector, $\|u(t)\|_2 \le 1$, one can use a polygonal approximation \citep{blackmore2012lossless}. The~considered task is for the ego spacecraft to travel from the initial condition $x_i = (-1,0,-1,0,0,0)$ to the target condition $x_f = (0,0,0,0,0,0)$ in the minimum time.

Assume we do not have the model \eqref{equ:31}. This may be due to uncertainty about the target's orbital rate~$\omega$ or uncertainty about the ego spacecraft's precise mass~$m_s$, the maximum thrust $T_{\max}$ its propulsion system produces, or thruster alignment. And assume we are only able to measure relative positions, i.e.,
\begin{equation}\label{equ:33}
y(t) = C x(t) = \begin{bmatrix} I_3 & 0_{3,3}
\end{bmatrix} x(t)
\end{equation}
We apply the approach developed in this paper to solve the minimum-time trajectory planning problem for the ego spacecraft in such a setting.

First, we build up a data-based model $(\mathcal{H}_L(u^d),\mathcal{H}_L(y^d))$ with $L = 40$ using a pair of input-output trajectories $(u^d,y^d)$ of length $M = 10,000$. It is checked that the persistent excitation of order $L + n = 46$ condition is satisfied by this trajectory. The lag of this system is $l_{\min} = 2$. Therefore, we choose $K_i = 2$ and $K_f = 2$ and consider the following initial condition for $({\bf u}_{[-2:-1]},{\bf y}_{[-2:-1]})$ and terminal condition for ${\bf y}_{[T:T+1]}$:
\begin{subequations}\label{equ:34}
\begin{align}
& {\bf u}_i \!=\! \text{col}\big((0,0,0),(0,0,0)\big),\, {\bf y}_i \!=\! \text{col}\big((-1,0,-1),(-1,0,-1)\big) \\
& Y_f = {\bf y}_f = \text{col}\big((0,0,0),(0,0,0)\big)
\end{align}
\end{subequations}
These conditions equivalently express the initial and terminal conditions $x_i$ and~$x_f$ for the state vector~$x$. We estimate that the minimum time $T^*$ is in the range of $[T_0,T_1] = [100,140]$, and hence we choose $K = \lceil \frac{T_1+K_f}{L-K_i} \rceil = 4$. In the LP formulation \eqref{equ:27}, we use $\theta = 2$, which is shown to be sufficiently large to yield a~minimum-time trajectory solution.

To validate the result obtained by our approach, we also~implement a (state-space) model-based mixed-integer programming (MIP) approach for minimum-time trajectory optimization, which is modified from the approach of \cite{carvallo1990milp}. The MIP formulation for the considered spacecraft relative motion planning problem is given as:
\begin{subequations}\label{equ:35} 
\begin{align}
\min_{u(\cdot), x(\cdot), \delta(\cdot)}\!\! &\quad \sum_{t = T_0}^{T_1} t\, \delta(t) \label{equ:351} \\
\text{s.t.}\quad &\text{dynamic constraints: } \label{equ:352} \\
& x(t+1) = Ax(t) + Bu(t),\,\, t = 0, ..., T_1-1 \nonumber \\
&\text{initial condition: }\, x(0) = x_i \label{equ:353} \\
&\text{path constraints: } \nonumber \\
& -1_m \le u(t) \le 1_m,\,\, t = 0, ..., T_1-1 \label{equ:354} \\
&\text{terminal condition: } \nonumber \\
& x(t) - x_f \le (1-\delta(t)) W 1_n \nonumber \\
& x_f - x(t) \le (1-\delta(t)) W 1_n \label{equ:355}  \\
& \delta(t) \in \{0,1\},\,\,\,\, t = T_0, ..., T_1 \nonumber \\
& \sum_{t = T_0}^{T_1} \delta(t) = 1 \label{equ:356} 
\end{align}
\end{subequations}
where $n = 6$, $m = 3$, $\delta(t) \in \{0,1\}$, $t = T_0, ..., T_1$, are indicator variables, $W > 0$ is a large constant, the constraints in \eqref{equ:355} mean that the state reaches the target condition $x_f$ at time $t$ if $\delta(t) = 1$, the constraint in \eqref{equ:356} makes sure that a feasible trajectory must reach $x_f$ at some $t \in [T_0,T_1]$, and the goal represented by the cost function in \eqref{equ:351} is to minimize the time to reach $x_f$.

\begin{figure}[htbp!]
\begin{center}
\begin{picture}(240.0, 180.0)
\put(0,0){\epsfig{file=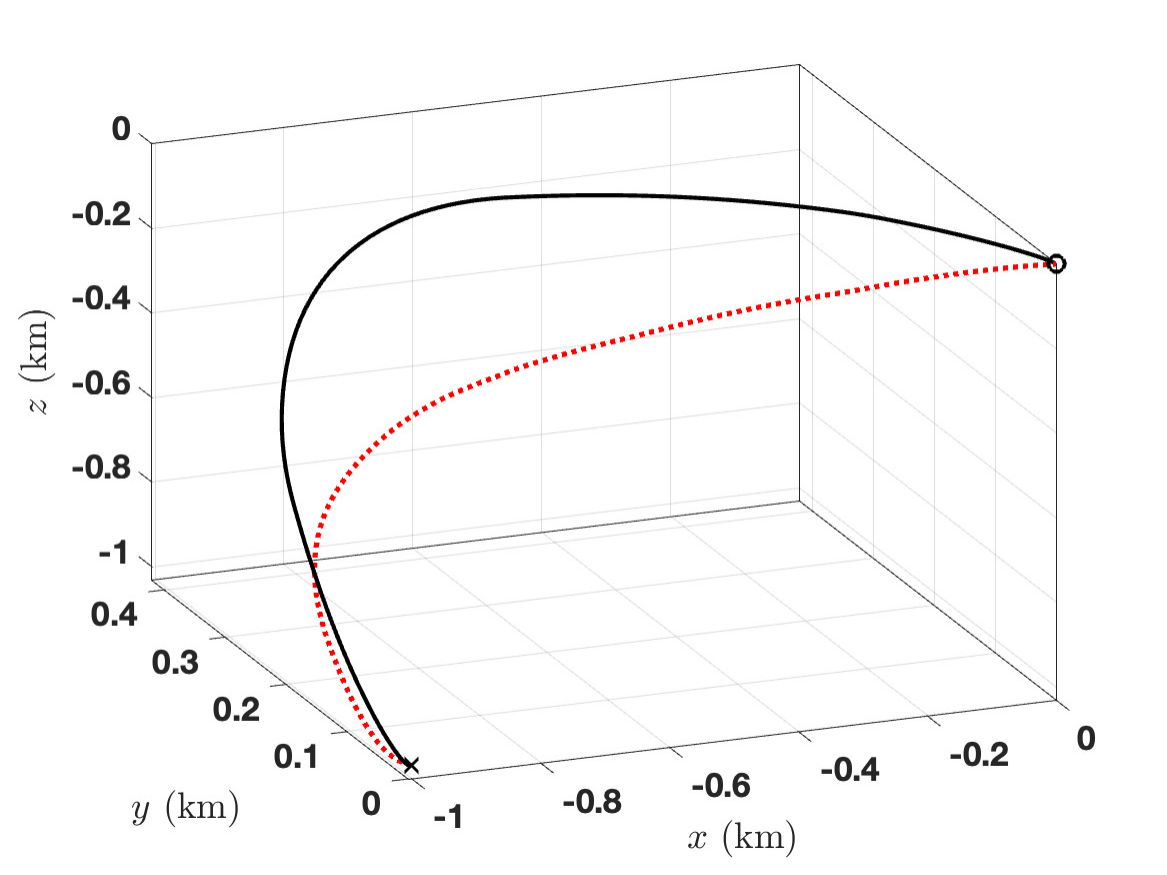,height=.36\textwidth}}
\end{picture}
\end{center}
      \caption{Spacecraft position trajectories from our LP approach~(black solid) versus from the model-based MIP approach~(red dotted).}
      \label{fig:2}
\end{figure}

\begin{figure}[htbp!]
\begin{center}
\begin{picture}(240.0, 190.0)
\put(-10,0){\epsfig{file=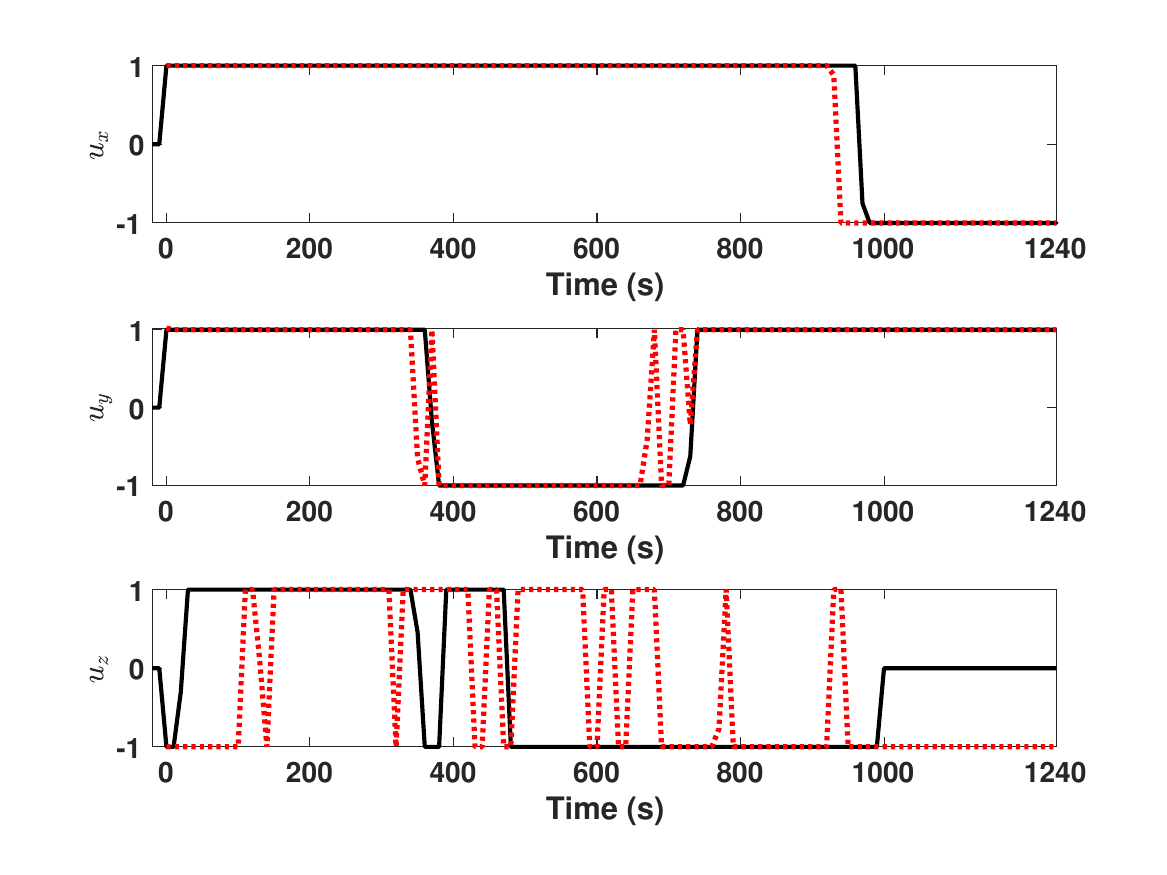,height=.38\textwidth}}
\end{picture}
\end{center}
      \caption{Thrust level vector time histories from our LP approach~(black solid) versus from the model-based MIP approach~(red dotted).}
      \label{fig:3}
\end{figure}

\begin{figure}[htbp!]
\begin{center}
\begin{picture}(240.0, 180.0)
\put(0,0){\epsfig{file=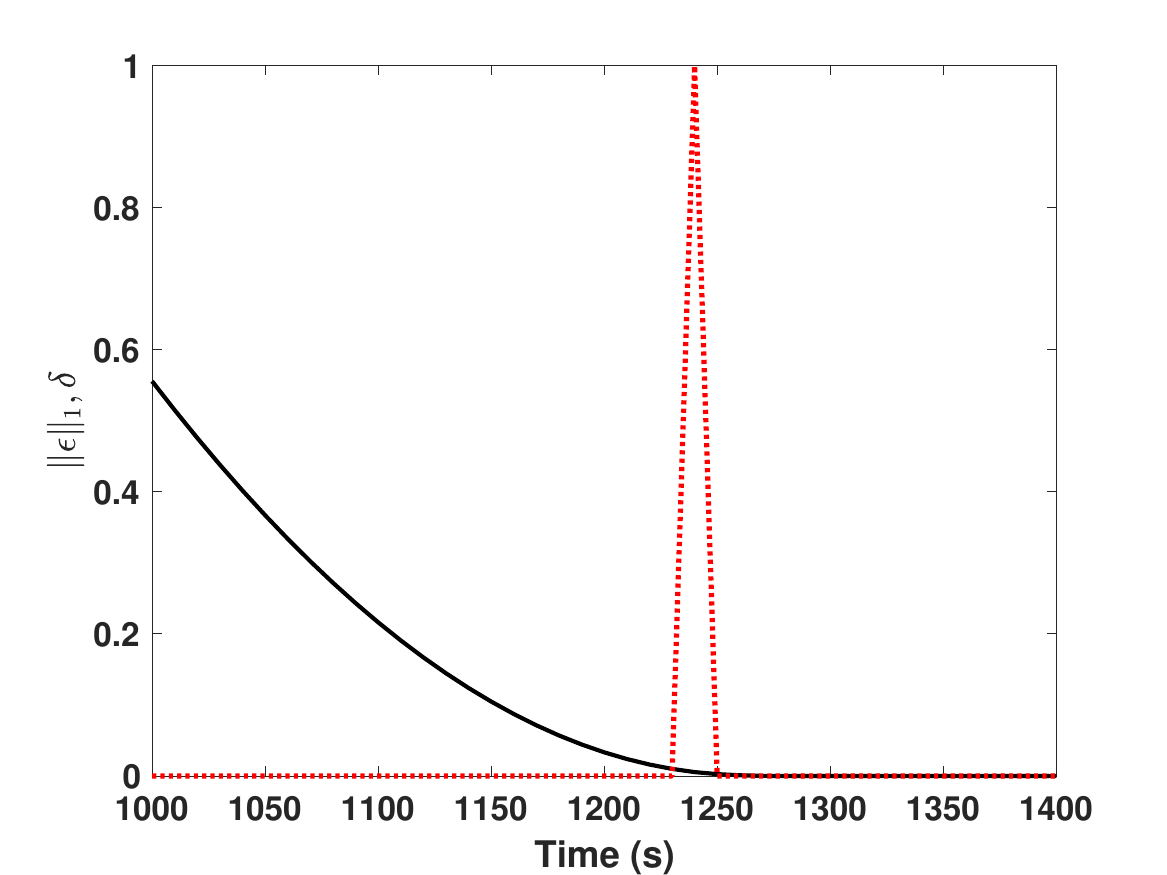,height=.36\textwidth}}
\end{picture}
\end{center}
      \caption{Slack variable values: $\varepsilon_t$ in our LP approach (black solid) versus $\delta(t)$ in the model-based MIP approach (red dotted).}
      \label{fig:4}
\end{figure}

Figs.~\ref{fig:2}--\ref{fig:4} illustrate the trajectory solutions from our LP approach based on data-based model (black solid curves) and the MIP approach based on state-space model (red dotted curves). Fig.~\ref{fig:2} shows the ego spacecraft relative position trajectories, and Fig.~\ref{fig:3} shows the thrust level vector time histories. It can be seen that the trajectories from the two approaches are different, especially for the {\sf z}-direction. However, the two trajectories have the same time-of-flight, $TOF = T^* \times \Delta t = 124 \times 10 = 1240$~(s), where $T^* = 124$ is the minimum (discrete) time obtained by both approaches and $\Delta t = 10$ (s) is the sampling period, i.e., the two trajectories are both minimum-time trajectories. It is well-known that in a discrete-time setting minimum-time trajectories are typically not unique. While the MIP approach~\eqref{equ:35} returns an arbitrary minimum-time solution, our~LP approach \eqref{equ:27} returns a minimum-time solution that also minimizes the secondary objective function $J_0 = \sum_{t = T_0}^{T^*-1} \theta^{t-T_0} \|{\bf y}_{[t:t+1]} - {\bf y}_f\|_1$ (roughly, the cumulative distance from the target condition). Correspondingly, it can be seen from Fig.~\ref{fig:3} that the thrust vector profile obtained by our LP approach is smoother than that from the MIP approach\footnote{To avoid obtaining a non-smooth control profile due to the non-uniqueness of minimum-time trajectories, one can add a small regularization term to the cost function \citep{carvallo1990milp}. In our LP approach, the secondary objective function $J_0 = \sum_{t = T_0}^{T^*-1} \theta^{t-T_0} \|{\bf y}_{[t:t+1]} - {\bf y}_f\|_1$ plays the role of such a regularization term.}. Fig.~\ref{fig:4} shows the values of the slack variables: $\varepsilon_t$ in our LP approach \eqref{equ:27} and $\delta(t)$ in the MIP approach \eqref{equ:35}. At $t = T^* = 124$ (corresponding to continuous time $T^* \times \Delta t = 1240$~(s)), $\varepsilon_t$~converges to zero (according to the criterion $\|\varepsilon_t\|_1 < 10^{-3}$) and $\delta(t)$ equals one, indicating that both approaches capture the minimum time $T^* = 124$.

In terms of the computations, the difference between the two approaches is significant: We solve the problems in~the {\sf MATLAB} environment running on a {\sf Windows 10 Enterprise OS} desktop with {\sf Intel Xeon Gold 2.30 GHz} processor and $32$ GB of RAM. It takes the LP approach $76$~milliseconds (averaged over $10$ experiments) to find a minimum-time solution (after elimination of the auxiliary variables $\zeta_k$ using the approach in {\it Remark~3} and solved with {\sf MATLAB linprog} function and default settings); while it takes the MIP approach $4.43$ seconds to find one (solved with {\sf MATLAB intlinprog} function and default settings) -- our LP approach based on data-based model is approximately $58$ times faster than the MIP approach based on state-space model.

\section{Conclusions}\label{sec:6}

In this paper, we developed an LP-based approach to minimum-time trajectory optimization using input-output data-based models. The approach was based on an effective integration of non-parametric data-based models for trajectory prediction and a continuous optimization formulation using an exponential weighting scheme for minimum-time trajectory planning. We~proved that minimum-time trajectories could be obtained with the approach if the weight parameter was chosen to be sufficiently high. We validated the approach and illustrated its application with an aerospace example. Future work includes integrating the approach into a model predictive control framework to achieve repeated replanning and closed-loop control, where real-time data may be used to update the model, and extending the approach to handle noisy data and nonlinear systems.

% \section*{Acknowledgments}

\bibliographystyle{ifacconf-harvard}
\bibliography{ref}

\end{document}